\documentclass[prb,twocolumn,amsmath,amssymb,floatfix]{revtex4}
\usepackage{graphicx}
\usepackage{dcolumn} 
\usepackage{bm}
\usepackage{amsmath}
\usepackage{latexsym}
\usepackage{amsfonts}
\usepackage{dcolumn}
\usepackage{color}
\usepackage{amssymb}
\usepackage[rightcaption]{sidecap}
\newcommand{\be}{\begin{equation}}
\newcommand{\ee}{\end{equation}}
\newcommand{\bea}{\begin{eqnarray}}
\newcommand{\eea}{\end{eqnarray}}
\newcommand{\nn} {\nonumber}
\renewcommand{\vr} {{\bf r}}

\def\g{\gamma}

\def\d{\delta}
\def\D{\Delta}

\def\ve{\varepsilon}

\def\S{\Sigma}

\def\vf{\varphi}

\def\w{\omega}

\def\bra{\langle}
\def\ket{\rangle}

\def\xc{{\rm xc}}
\def\x{{\rm x}}

\begin{document}
\title{Effect of discontinuities in Kohn-Sham-based chemical reactivity theory}
\author{Maria Hellgren}
\author{E. K. U. Gross}
\affiliation{Max Planck Institute of Microstructure Physics, Weinberg 2, 06120 Halle, Germany}  
\date{\today}
\begin{abstract}
We provide a new derivation of a formula for the Fukui function of density-functional chemical reactivity theory which incorporates the discontinuities in the Kohn-Sham reference system. Orbital relaxations are described in terms of the exchange-correlation (XC) kernel, i.e., the derivative of the XC potential with respect to the density and it is shown that in order to correctly measure the reactivity toward a nucleophilic reagent a discontinuity of the XC kernel has to be taken into account. The importance of this finding is illustrated in model molecular systems.  
\end{abstract}
\maketitle
\section{Introduction}
Predicting how molecules respond to external perturbations is an important subject in theoretical chemistry. On a fundamental level this entails a very 
difficult problem as molecules are composed of interacting electrons and nuclei which require a solution to the many-body Schr\"odinger equation. In an attempt to simplify the description a set of reactivity descriptors and associate empirical equalization principles have been formulated, constituting what is known as chemical reactivity theory (CRT).\cite{fukjcp,fuksci,parryang84,ccp95,ib97,ap99,payaayle,amr02,gpl03,gp08,ap08,gi09,parrbook} An approach which naturally combines with CRT is density functional theory (DFT),\cite{hk,vBscripta} which is an exact framework for treating the electron-electron interaction in terms of only the electronic density. Normally, the Kohn-Sham (KS) formulation\cite{KS} is used in which the density is calculated from a fictitious system of non-interacting electrons moving in an effective local multiplicative potential - the KS potential. 
It is well known that an independent-particle description using a local potential sometimes introduces singularities in the KS quantities and that many physical properties are crucially dependent on this singular behavior. A classic example is the dissociation of closed-shell molecules composed of open-shell atoms.\cite{perde,ulfca} In order to dissociate with a correct integer number of electrons on each atom a discontinuous positive step in the exchange-correlation (XC) part of the KS potential has to develop over the atom with the larger ionization potential. Important progress was made when it was understood that this feature can be related to a derivative discontinuity of the total energy as a function of particle number.\cite{pplb82} In order to make this identification DFT was generalized to ensembles allowing for fractional charges. 

In CRT the use of non-integer number of particles is at the basis of all definitions of reactivity indices since the reaction probability is mostly measured in terms of a sensitivity to a global or local change of particle number. In KS-based CRT one therefore expects singularities as those discussed above to become important. The aim of this paper is to show an example which points out this fact, namely in the determination of the Fukui function, a local reactivity index. In this case, the quantity having the crucial discontinuity is the XC kernel defined as the functional derivative of the XC potential with respect to the density. 
The XC kernel is the quantity that accounts for the effects of orbital relaxation, which may produce a large difference when predicting the reactivity of certain molecules.\cite{ba05} In this paper, we will show how a discontinuity of the XC kernel enters when determining the Fukui function and that this quantity is in fact what gives the largest contribution in describing the reaction toward a nucleophilic reagent. 

The paper is organized as follows. In Sec. II we start by deriving an expression for the Fukui function in terms of the XC kernel. In Sec. III we discuss discontinuities in DFT with a particular emphasis on the discontinuities of the XC kernel and show their importance for the formula derived in Sec. II. A numerical investigation in terms of two-electron model molecular systems in the exact-exchange (EXX) approximation is given in Sec. VI. Finally, we give our conclusions in Sec. V. 
\section{Fukui function}
In order to define quantities such as the chemical potential or Fukui functions which involve derivatives with respect to the number of particles the theory must be generalized to systems which involve fractional charges. This implies an ensemble description in terms of states with different electron numbers. For an average number of electrons $N=N_0+\w$, where $N_0$ is an integer and $0<\w<1$ Perdew et al.\cite{pplb82} proposed an ensemble of the form 
\be
\hat{\g}^{>}= (1-\w)|\Psi_{N_0}\ket\bra\Psi_{N_0}|+\w|\Psi_{N_0+1}\ket\bra\Psi_{N_0+1}|,
\label{glarge}
\ee 
where $\Psi_{k}$ is the ground state wave function of $k$ particles. Similarly, for $N=N_0-1+\w$ we can define
\be
\hat{\g}^{<}= (1-\w)|\Psi_{N_0-1}\ket\bra\Psi_{N_0-1}|+\w|\Psi_{N_0}\ket\bra\Psi_{N_0}|.
\label{gsmall}
\ee 
Using these ensembles any derivative with respect to $N$ is equal to the derivative with respect to $\w$.

The ensemble ground-state energy $E(N)$ will consist of straight line segments between the values at the integers. Hence, the chemical potential $\mu=\partial E/\partial N$, i.e., the slope, on the $-/+$ side of $E(N_0)$ is equal to the negative of the ionization energy and affinity  (I/A), respectively. 

The Fukui function is defined as 
\bea
f(\vr)=\frac{\partial n(\vr)}{\partial N}=\frac{\d \mu}{\d w(\vr)},
\eea
where $n(\vr)$ is the electron density and $w(\vr)$ the external potential. These two definitions are equal which is easily seen from the identity $\d E/\d w(\vr)=n(\vr)$. Using the first definition we immediately see that the Fukui function is constant (independent of $N$) between integers. The limits $N\to N_0^{\pm}$ is, however, usually what is of interest and we therefore write
\bea
f^+(\vr)&=&n_{N_0+1}(\vr)-n_{N_0}(\vr)\\
f^-(\vr)&=&n_{N_0}(\vr)-n_{N_0-1}(\vr),
\eea
where here and in the following the $+/-$ sign refers to these different limits. These results are a direct consequence of using the above defined ensembles, where the species is assumed to be completely independent of its environment. To address this aspect, the concept of chemical-context dependent reactivity descriptors has been proposed.\cite{cowassfound} 

So far we have not used DFT but it appears to be a natural framework for calculating reactivity indices like the Fukui function which is defined in terms of the density only. The extension of DFT to fractional charges\cite{pplb82} allows us to formally take derivatives with respect to $N$. The density is usually calculated within the KS framework which assumes the ensemble density to be non-interacting ensemble $v$-representable. The ensemble KS potential, $V_s$, has been the subject of several investigations all showing that the XC part $v_\xc(\vr)$ must have a discontinuous behavior as a function of $N$,\cite{gorisav,sagper} a topic we will elaborate further on in the next section.

In order to derive expressions for the Fukui function in terms of KS quantities we start by writing the density corresponding to Eqs. (\ref{glarge}-\ref{gsmall}) in terms of KS orbitals
\bea
n^{>}(\vr)&=&\sum_k^{N_0}|\vf^{\w}_k(\vr)|^2+\w|\vf^{\w}_{N_0+1}(\vr)|^2\\
n^{<}(\vr)&=&\sum_k^{N_0-1}|\vf^{\w}_k(\vr)|^2+\w|\vf^{\w}_{N_0}(\vr)|^2.
\eea
We notice that the orbitals depend on $\w$ (or $N$) since the ensemble KS potential will be different for every value of $\w$. Keeping this in mind we can take the derivative with respect to $N$ and determine the Fukui functions\cite{ayb}
\bea
f^+(\vr)&=&|\vf_{N_0+1}(\vr)|^2+\sum_k^{N_0}\frac{\partial |\vf_k(\vr)|^2}{\partial N}\\
f^-(\vr)&=&|\vf_{N_0}(\vr)|^2+\sum_k^{N_0}\frac{\partial|\vf_k(\vr)|^2}{\partial N}.
\eea
The superscript $\w$ on the orbitals is now dropped since the limit $N\to N_0^\pm$ has been taken. The orbitals are continuous with respect to $N$ and are therefore unambiguously determined by the $N_0$-system. The frontier molecular orbital (FMO) approximation corresponds to ignoring orbital relaxations, i.e., $f^+(\vr)\approx |\vf_{N_0+1}(\vr)|^2$ and $f^-(\vr)\approx |\vf_{N_0}(\vr)|^2$. In the formulation by Parr et al.\cite{parrbook} orbital relaxations due to Coulomb interactions can be taken into account via the second terms on the right hand side of Eqs. (8-9). The differentiation of the orbitals can easily be performed using the chain rule 
\be
\frac{\partial \vf_k(\vr)}{\partial N}=\int d\vr' \frac{\delta \vf_k(\vr)}{\delta V_s(\vr')}\frac{\partial V_s(\vr')}{\partial N}.
\ee
A variation in the number of particles will induce a variation in the KS potential via the density
\bea
\frac{\partial V_s(\vr')}{\partial N}&=&\int d\vr''\frac{\delta [v_{\rm H}(\vr')+v_{\xc}(\vr')]}{\delta n(\vr'')}f(\vr'')\nn\\
&=&\int d\vr''[v(\vr',\vr'')+f_\xc(\vr',\vr'')]f(\vr''),
\eea
where $v_{\rm H}(\vr)=\int d\vr' v(\vr,\vr')n(\vr')$ is the Hartree potential, $v$ is the bare Coulomb interaction, or the Hartree kernel, and $f_\xc$ is the XC kernel. It is then easy to see
\bea
f^\pm(\vr)&=&f^\pm_0(\vr)\nn\\
&&+\int d\vr' d\vr''\,\chi_s(\vr,\vr')f^{\pm}_{\rm H\xc}(\vr',\vr'')f^\pm(\vr''),
\label{fukuichi0}
\eea
where $\chi_s=\delta n/\delta V_s$ is the KS density response function, $f_{\rm H\xc} = v+f_\xc$ and $f^\pm_0$ is the Fukui function in the FMO approximation defined above. Here we have been careful in taking the limit $N\to N_0^\pm$ of the XC kernel since it has been recently shown that $f_\xc$ has discontinuities.\cite{hg11} Defining the matrix 
\bea
K^{\pm}(\vr,\vr')=\d(\vr,\vr')-\int d\vr''\,\chi_s(\vr,\vr'')f^{\pm}_{\rm H\xc}(\vr'',\vr')
\eea
we can recast Eq. (\ref{fukuichi0}) into
\bea
f^\pm(\vr)&=&\int d\vr'\,[K^{\pm}(\vr,\vr')]^{-1}f^\pm_0(\vr').
\label{chisf}
\eea
This formula was obtained by Cohen et al.\cite{copiku1,copiku2} apart from the fact that we here allow the kernel to have discontinuities. That these play a dominant role when calculating the Fukui function $f^+$ will be 
shown in Sec. IV. 

We will now show an alternative derivation of Eq. (\ref{chisf}) based on the equivalent definition of the Fukui function as the functional derivative of the chemical potential with respect to the external potential.\cite{apbg07,spag07,fspag08} We begin by evaluating the chemical potential and show its equivalence with the highest occupied eigenvalue of the KS system. The ground-state energy is a functional of the density and can be written as
\bea
E[n]&=&T_s[n]+\frac{1}{2}\int d\vr d\vr'  n(\vr)v(\vr,\vr')n(\vr')\nn\\
&&+\int d\vr\, w(\vr) n(\vr)+E_\xc[n]
\eea
where $T_s$ is the non-interacting kinetic energy functional and $E_\xc$ is the XC energy. Taking the derivative with respect to the number of particles we write
\bea 
\frac{\partial E}{\partial N}=\frac{\partial T_s}{\partial N}+\int d\vr\left [w(\vr)+v_{\rm H}(\vr)\right ] f(\vr)+\frac{\partial E_\xc}{\partial N},
\eea
where we have identified the Fukui function. The derivative of the kinetic energy can easily be evaluated once written in terms of occupied KS eigenvalues $\ve_k$. Let us first focus on an ensemble of the form of Eq. (\ref{glarge}), i.e., with $N=N_0+ \w$. We can then write
\begin{widetext}
\bea
\frac{\partial E^>}{\partial N}=\frac{\partial}{\partial N}\left [\sum_k^{N_0}\ve_k+\w\ve_{N_0+1}-\int d\vr\left [w(\vr)+v_{\rm H}(\vr)+v_\xc(\vr)\right] n(\vr)\right ]+\int d\vr\left [w(\vr)+v_{\rm H}(\vr) \right]f(\vr)+\frac{\partial E_\xc}{\partial N}.
\eea
\end{widetext}
The eigenvalues are functions of $N$ via the KS potential and we find after a few manipulations
\bea
\frac{\partial E^>}{\partial N}=\ve_{N_0+1}-\int d\vr\, v_\xc(\vr) f(\vr)+\frac{\partial E_\xc}{\partial N}=\ve_{N_0+1},
\label{chemeig1}
\eea
where we have used the identity
\bea
\frac{\partial E_\xc}{\partial N}=\int d\vr \,v_\xc(\vr) f(\vr)
\eea
and the definition of the XC potential $v_\xc=\d E_\xc/\d n$. The same steps can be performed for the ensemble in Eq. (\ref{gsmall}) and we find similarly 
\bea
\frac{\partial E^<}{\partial N}=\ve_{N_0}.
\label{chemeig2}
\eea
Eqs. (\ref{chemeig1}) and (\ref{chemeig2}) thus prove that the highest occupied eigenvalue must be equal to the chemical potential\cite{pl97} and should therefore not change with $N$. Many problems with existing functionals are related to a lack of this straight-line behavior when extended to fractional charges.\cite{cmsy08,mscy09} In the limit $N\to N_0^\pm$ we have
\bea
\left.\frac{\partial E}{\partial N}\right|_+=\ve_{N_0+1}^+,\,\,\,\, \,\,\,\left.\frac{\partial E}{\partial N}\right|_-=\ve_{N_0}^-.
\label{deriveplus}
\eea
In this limit $\ve_{N_0+1}^+=\ve^+_{\rm LUMO}$, i.e., the lowest unoccupied KS orbital obtained from the KS potential in the limit $N\to N_0^+$ ($V_s^+$). In the same way $\ve_{N_0}^-=\ve^-_{\rm HOMO}$, i.e., the highest occupied KS orbital obtained from $V_s^-$. These different limits are important to keep since a constant shift (or discontinuity) in $v_\xc$ has been shown to occur as an integer is crossed. The discontinuity in $v_\xc$ is in general positive, shifting the KS affinity $A_s=\ve^-_{\rm LUMO}$ to the true affinity $A=\ve^+_{\rm LUMO}$ in order to obey the relation in Eq. (\ref{chemeig1}). Before continuing the discussion on discontinuities (Sec. III) we will determine the Fukui function from
\bea
f(\vr)=\frac{\d \mu}{\d w(\vr)}=\frac{\d}{\d w(\vr)}\frac{\partial E}{\partial N}=\frac{\d \ve_{N_0/N_0+1}}{\d w(\vr)}.
\eea
The variation of an KS eigenvalue with respect to the external potential can straightforwardly be obtained from first order perturbation theory. Taking into account that varying $w$ will induce a variation of $v_{\rm H}$ and $v_\xc$ via the density we find 
\bea
f^\pm(\vr)&=&f^\pm_0(\vr)\nonumber\\
&&+\int d\vr' d\vr''\chi(\vr,\vr'')f^\pm_{\rm H\xc}(\vr'',\vr')f^\pm_0(\vr').
\label{chifuk}
\eea
Using the relation
\bea
\chi=\chi_s+\chi_s [v+f_\xc]\chi
\label{hardy}
\eea
it is easy to see that Eqs. (\ref{chifuk}) and (\ref{chisf}) are equivalent.
In the following we will prefer the use of Eq. (\ref{chifuk}) since any discontinuity of $f_\xc$ enters linearly. 
\section{Discontinuities in DFT}
It has been shown that the ground-state energy exhibits derivative discontinuities at integer particle number.\cite{pplb82} For non-interacting electrons in a system with discrete energy levels this happens only when $N$ crosses an integer for which a new orbital with a different eigenvalue starts to be occupied. In this case, the effect is mainly due to the kinetic energy. In an interacting system with discrete energy levels a large part of the derivative discontinuity will be contained in $E_\xc$ and the discontinuity will even show up when $N$ crosses an integer lying within a shell with the same eigenvalue. 

A derivative discontinuity in $E_\xc$ gives rise to a discontinuity in $v_\xc$ in the form of a constant shift $\D_\xc$. The same discontinuity also leads to discontinuities in the XC kernel $f_\xc$ which are of more complex nature than those of $v_\xc$. In this section we briefly review the discontinuities of $v_\xc$ and $f_\xc$. 
\subsection{XC potential}
Any discontinuity of $v_\xc$ is related to a derivative discontinuity in $E_\xc[n[w,N]]$. Using the chain rule we can write
\bea
\frac{\partial E_{\xc}}{\partial N}=\int d \vr\,  v_{\xc}(\vr)\frac{\partial n(\vr)}{\partial N}.
\label{forstaderivN1}
\eea
This identity has to hold true for any value of $N$ and we can use it to formally write down the value of the discontinuous shift $\D_\xc$ at $N_0$. 
For $N>N_0$ we write $v^>_{\xc}(\vr)=v^-_{\xc}(\vr)+\D_\xc(\vr)$ and insert this expression in Eq. (\ref{forstaderivN1}) 
\bea
\frac{\partial E^>_{\xc}}{\partial N}=\int d \vr\,\,  \left[v^-_{\xc}(\vr)+\D_\xc(\vr)\right]f(\vr).
\label{forstaderivN}
\eea
Now we take the limit $N\to N_0^+$ and after a rearrangement we find
\bea
\D_\xc=\left.\frac{\partial E_{\xc}}{\partial N}\right |_+-\int d \vr\, v^-_{\xc}(\vr)f^+(\vr),
\label{forstaderivN2}
\eea
where we have used the fact that the Fukui function integrates to unity. If $\partial E_{\xc}/\partial N$ is discontinuous at $N_0$, $\D_\xc$ will be finite.\cite{fndisc} As an example we can use the case where $E_\xc$ is a functional of the KS Green function $G_s$ and $\d E_\xc/\d G_s=-i\S_\xc$, where $\S_\xc$ is called the self-energy. Using Eq. (\ref{forstaderivN2}) we find the celebrated MBPT formula for the discontinuity\cite{perde}
\bea
\D_\xc&=&\int d\vr d\vr' \vf_{\rm L}(\vr)\left[\S_\xc(\vr,\vr',\ve^+_{\rm LUMO})\right.\nn\\
&&\,\,\,\,\,\,\left.-v^-_\xc(\vr')\d(\vr-\vr')\right]\vf_{\rm L}(\vr'),
\label{perdew}
\eea
where $\vf_{\rm L}$ is the LUMO orbital. In the next section we will use this formula in the EXX approximation for which the self-energy corresponds to the Hartree-Fock (HF) self-energy but evaluated with KS orbitals. 

A discontinuous shift in the XC potential implies also a shift in the eigenvalues with the same magnitude. We can thus write Eq. (\ref{deriveplus}) in the previous section as
\bea
A=\left.\frac{\partial E}{\partial N}\right|_+=\ve_{\rm LUMO}^-+\D_\xc,
\eea
that is, the affinity is equal to the KS affinity plus the discontinuity, a well-known result.\cite{perde} 
\subsection{XC kernel}
In order to determine the discontinuities of the XC kernel we start by noting that for particle number conserving variations of the density $f_\xc$ is only defined up to the sum of two arbitrary functions $g_\xc(\vr)+g_\xc(\vr')$. This observation follows immediately after inspecting the definition of $f_\xc$
\bea
\d E_\xc=\int d\vr'd\vr f_\xc(\vr,\vr')\d n(\vr')\d n(\vr).
\eea
When we allow for non particle conserving density variations the arbitrariness disappears but leaves the functional discontinuous. In the case of the XC kernel these discontinuities will have to take the form 
\bea
f^+_\xc(\vr',\vr)-f^-_\xc(\vr',\vr)=g_\xc(\vr)+g_\xc(\vr').
\eea
In order to determine $g_\xc$ with the same procedure used for the XC potential we study the quantity
\be
\frac{\d}{\d w(\vr)}\frac{\partial E_\xc}{\partial N}.
\ee
Varying Eq. (\ref{forstaderivN1}) with respect to the external potential allows us to write this quantity in terms of the XC kernel. The kernel can then be written as $f^>_\xc(\vr',\vr)=f^-_\xc(\vr',\vr)+g_\xc(\vr',\vr)$. Taking the limit $N\to N^+_0$ implies $g_\xc(\vr',\vr)\to g_\xc(\vr)+g_\xc(\vr')$ and we arrive at
\begin{widetext}
\bea
\int d \vr \,\chi(\vr_1,\vr) g_\xc(\vr)=\left.\frac{\d}{\d w(\vr)}\frac{\partial E_\xc}{\partial N}\right|_+
-\!\int\! d \vr d \vr'\,\chi(\vr_1,\vr) f^-_{\xc}(\vr,\vr')f^+(\vr')-\!\int \!d \vr \,
v^+_\xc(\vr)\frac{\d f^+(\vr)}{\d w(\vr_1)}.
\label{muxcdeltan0}
\eea
\end{widetext}
From this equation $g_\xc$ is only determined up to constant. This constant is, however, easily fixed by considering the the second derivative of $E_\xc$ with respect to $N$
\bea
 2\int d \vr \,f^+(\vr)g_\xc(\vr)&=&\left.\frac{\partial^2 E_\xc}{\partial N^2}\right|_+-\int  d \vr' \,v^+_\xc(\vr')\frac{\partial f^+(\vr')}{\partial N}
 \nonumber \\
&&\!\!\!\!\!\!\!\!\!\!\! -\int d \vr d\vr' \, f^+(\vr)f^-_{\xc}(\vr,\vr')f^+(\vr'),
\label{muxcdeltandeltan}
\eea
yielding a condition to be imposed on Eq. (\ref{muxcdeltan0}). The function $g_\xc$ obtained via Eqs. (\ref{muxcdeltan0}-\ref{muxcdeltandeltan}) was recently analyzed in Ref. \onlinecite{hg11} showing a diverging behavior of the form 
\be
g_\xc(\vr)\sim \frac{|\vf_{N_0+1}(\vr)|^2}{n(\vr)}\sim e^{2(\sqrt{2I}-\sqrt{2A_s})\, r},
\ee
as $r\to \infty$. 
The diverging behavior can be deduced by performing a common denominator approximation to Eq. (\ref{muxcdeltan0}).\cite{kli} That this is indeed a reliable approximation to Eq. (\ref{muxcdeltan0}) was shown in Ref. \onlinecite{hg11}. We can now go back to Eq. (23) and add the appropriate term to $f^+(\vr)$ arising from the discontinuity of $f_\xc$. We find
\begin{widetext}
\bea
f^+(\vr)=f_0^+(\vr)+\int d\vr'' d\vr'\chi(\vr,\vr')[f^-_\xc(\vr',\vr'')+v(\vr',\vr'')]f_0^+(\vr'')+\int d\vr '\chi(\vr,\vr')g_\xc(\vr'),
\label{fukder}
\eea
\end{widetext}
or 
\bea
f^+(\vr)=f_s^+(\vr)+\int d\vr' \chi(\vr,\vr')g_\xc(\vr'),
\label{fana}
\eea
which defines $f_s^+(\vr)$ as the Fukui function when the discontinuity is not taken into account.
\section{EXX approximation}
In order to quantify the relative importance of the two different contributions to $f^+(\vr)$ ($f_\xc^-$ and $g_\xc$) we have performed a numerical study on model 1D molecular systems in the EXX approximation. The EXX functional is known to contain the derivative discontinuity at even integers when defined on densities corresponding to spin-compensated  ensembles composed of states with different number of particles. 

The EXX energy functional is an implicit functional of the density given by\cite{gkkg}
\bea
E_\x&=&-\frac{1}{2}\int\! d\vr d\vr' \g(\vr,\vr')v(\vr,\vr')\g(\vr,\vr'),
\eea 
where the density matrix is given by $\g(\vr,\vr')=\sum_{k}^{\rm occ}\vf_k(\vr)\vf_k(\vr')$ after spin-summations have been performed. The corresponding EXX potential $v_\x=\d E_\x/\d n$ can be evaluated as
\be
\frac{\d E_\x}{\d V_s(\vr)}=\int\!d\vr' \frac{\d E_\x}{\d n(\vr')}\frac{\d n(\vr')}{\d V_s(\vr)}.
\label{lss}
\ee
Since $E_\x[\g]$ is an explicit functional of the density matrix the derivative $\d E_\x/\d V_s(\vr)$ is most conveniently evaluated using the chain-rule 
\bea
\frac{\d E_\x}{\d V_s(\vr)}&=&\int d\vr_1d\vr_2\frac{\d E_\x}{\d \g(\vr_2,\vr_1)}\frac{\d \g(\vr_2,\vr_1)}{\d V_s(\vr)}.
\eea
The potential is only determined up to constant by Eq. (\ref{lss}) but this constant may be fixed using Eq. (\ref{forstaderivN1}). The discontinuity in the EXX potential\cite{kli} is given by Eq. (\ref{perdew})
\bea
\D_\x&=&\int d\vr d\vr' \vf_{\rm L}(\vr)\left[-\g(\vr,\vr')v(\vr,\vr')\right.\nn\\
&&\,\,\,\,\,\,\left.-v^-_\xc(\vr')\d(\vr-\vr')\right]\vf_{\rm L}(\vr'),
\label{perdew}
\eea

In the case of two electrons the EXX potential takes a particularly simple form being equal to minus half the Hartree potential $v_\x(\vr)=-1/2\int d\vr' v(\vr,\vr')n(\vr')$. The EXX kernel is then also easily evaluated resulting in $f_\x(\vr',\vr)=-1/2 v(\vr',\vr)$. The expressions for the potential and kernel are evaluated for $N\to 2^-$. In the limit $N\to 2^+$ we have to add the discontinuity $\D_\x$ to $v_\x$ and two functions $g_\x$ to the kernel 
\bea
f^+_\x(\vr',\vr)=-\frac{1}{2}v(\vr',\vr)+g_\x(\vr)+g_\x(\vr').
\eea
We will now use Eq. (\ref{muxcdeltan0}) to determine $g_\x$. Using the chain-rule 
\bea
\frac{\d}{\d w(\vr)}=\int d\vr'\frac{\d}{\d V_s(\vr')}\frac{\d V_s(\vr')}{\d w(\vr)}
\eea 
this equation can be written in terms of $\chi_s$ only and we find
\begin{widetext}
\bea
\int d \vr \,\chi_s(\vr_1,\vr) g_\x(\vr)=\left.\frac{\d }{\d V_s}\frac{\partial E_\x}{\partial N}\right|_++\frac{1}{2}\!\int\! d \vr d \vr'\,\chi_s(\vr_1,\vr) v(\vr,\vr')|\vf_{\rm L}(\vr')|^2-\!\int \!d \vr \,
v^+_\x(\vr)\frac{\d |\vf_{\rm L}(\vr)|^2}{\d V_s(\vr_1)}.
\label{muxcdeltan}
\eea
\end{widetext}
To calculate the Fukui function we notice that only the quantity $\chi g_\x$ is needed. The arbitrary constant can therefore be left undetermined. We also notice that 
\be
f_s^+(\vr_1)=f_0^+(\vr_1)+\frac{1}{2}\int d\vr d\vr'\chi(\vr_1,\vr')v(\vr',\vr)f_0^+(\vr).
\ee
In the following section we will solve these two-electron equations for model molecular systems to illustrate the importance of the discontinuity.
\begin{figure}[t]
\includegraphics[width=8.5cm, clip=true]{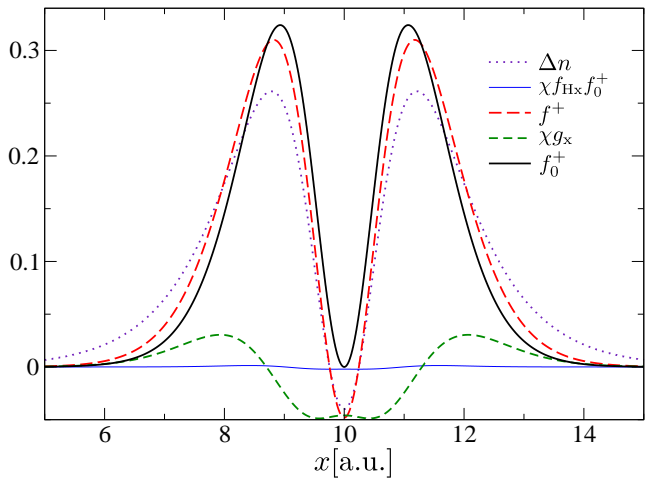}\\
\caption{The Fukui function of a 1D Be$^{2+}$ atom with the nucleus located at 10 a.u.. For comparison also the density difference $\D n(x)=n_{\rm N=3}(x)-n_{\rm N=2}(x)$ is plotted. The contribution from the discontinuity is seen to have the largest effect.}
\label{ediag}
\end{figure}
\begin{figure}[t]
\includegraphics[width=8.5cm, clip=true]{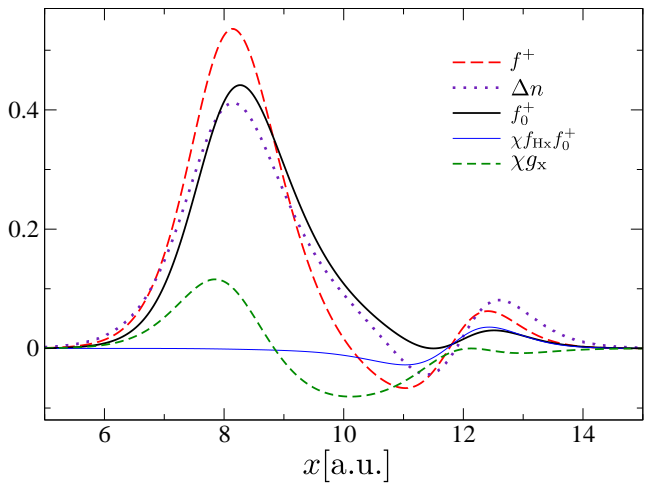}\\
\caption{The Fukui function of a 1D He$^{2+}$Be$^{2+}$ molecule with the nuclei located at 8 and 12 a.u.. For comparison also the density difference $\D n(x)=n_{\rm N=3}(x)-n_{\rm N=2}(x)$ is plotted. The contribution from the discontinuity is seen to have a large effect over the He atom.}
\label{ediag3}
\end{figure}
\subsection{Numerical results for model systems}
Our model systems consist of 1D molecules where the singular Coulomb interaction has been replaced with a soft-Coulomb interaction with softening parameter 1. The inter-particle Coulomb interaction is thus model by $1/\sqrt{(x_1-x_2)^2+1}$ and the external nuclear potentials by $Z/\sqrt{x_1^2+1}$, where $Z$ is the nuclear charge. This model has been used extensively in the literature\cite{maitrasc,helsc} to mimic real 3D diatomic molecules with results in  qualitative agreement, enough also for our purposes. The equations derived in the previous sections are now all evaluated consistently using the model potentials. 

The first system we study is a 1D ${\rm Be}^{2+}$ ($Z=4$) atom. In Fig. \ref{ediag} we show the Fukui function $f^+$ calculated using different approximations. The black solid curve corresponds to using the FMO approximation, i.e., $f^+(\vr)\approx f_0^+(\vr)=|\vf_L(\vr)|^2$. The dotted curve is obtained by calculating the densities of the $N=2$ and $N=3$ systems separately and then subtract them. For a functional with a linear behavior between the integers this approach should give the same result as calculating the derivative $f^+$. The derivative is given by the red long dashed curve and we see that although the qualitative features now agree there are still some discrepancies. The fact that $\D n$ and $f^+$ do not agree perfectly is related to the well known fact that the EXX functional is not linear between integers. The green dashed and blue solid thin curves shows the different contributions to the derivative (Eq. (\ref{fukder})). It is remarkable to see the effect of the discontinuity, which is very large compared to using only the $f_s$ approximation. The effect is to give a negative contribution at the nucleus and slow down the asymptotic decay. What is perhaps of more interest is the location of the peak which is seen to be shifted compared to the FMO peak, and hence becomes in good agreement with the $\D n$ result.
 
Next, we turn to a molecular system composed of 1D He$^{2+}$Be$^{2+}$ (Fig. \ref{ediag3}). What is particular about this system is that the HOMO and LUMO orbitals are spatially well separated. The HOMO is located at the Be site and the LUMO at the He site. The density response function contains only products of occupied and unoccupied orbitals, which decay exponentially with nuclear separation. In this case it is thus clear that the correction in $f_s$ can only be very small, unless $f_\x^-$ becomes very large, which is not the case since $f_\x^-=1/2v$. The blue solid thin curve in the figure confirms this fact. We have, however, seen that the discontinuity is a diverging function and can thus compensate for small excitation functions in the response function. Indeed, we find a large contribution to the Fukui function from the discontinuity. Also in this case we see that the peak position is improved compared to the FMO approximation. From an LDA-type of functional such improvement could not be achieved due to the lack of a diverging discontinuity. This further suggests that in these cases $f^+$ would be more accurately calculated by subtracting densities at different integer particle numbers. 

To test the consistency of our results we have also obtained the Fukui function from the numerical derivative, i.e., we have calculated 
\be
f^{Nder}(x)=\frac{n_{N=2+\D N}(x)-n_{N=2}(x)}{\D N}
\ee
for small values of $\D N$ using the ensemble of Eq. (\ref{gsmall}) in the EXX functional. When $\D N\to 0 $ this result should coincide with $f^+$ calculated from Eq. (\ref{fana}). In Fig. 3, we show that they, indeed, agree very well. 
\section{Conclusions}
In this paper we have derived equations for the Fukui function of DF CRT using a KS reference system. Our central result is that a discontinuity of the XC kernel enters when calculating the Fukui function for a nucleophilic attack ($f^+$). The importance of this result has been demonstrated in model molecular systems, where the effect of orbital relaxation has been shown to be almost entirely due to the discontinuity. From this we can conclude that in any system where orbital relaxations are important the discontinuity must be incorporated. These conclusions are of course based on model systems but they naturally carry over to real three dimensional atoms and molecules. An implementation of the full EXX functional for a molecule is, however, quite demanding and we therefore leave such investigation for future work. 

We would also like to point out that using the HF method to calculate Fukui functions\cite{bg05} in the FMO approximation as opposed to the KS method might yield very different results for $f^+$. In HF theory the orbitals are determined from a non-local potential, yielding very different virtual orbitals. If the HF LUMO orbital is closer to the true Fukui function is, however, hard to say.  

As a final remark we have in this paper only considered the Fukui function. A whole set of other reactivity descriptors exist and, in general, discontinuities will show up in the derivatives when using a KS reference system. In particular, we believe that for the calculation of the local hardness or hardness kernels\cite{bp87,ap208,crgt07} discontinuities will be an essential ingredient. 
\begin{figure}[t]
\includegraphics[width=8.5cm, clip=true]{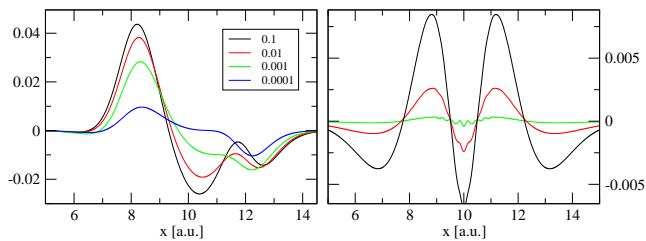}\\
\caption{The difference between the analytic ($f^+$) and the numerical $\left\{n(N=2+\D N)-n(N=2)\right\}/\D N$ derivatives at $\D N=0.1,0.01,0.001,0.0001$. To the left: He$^{2+}$Be$^{2+}$ and to the right: Be$^{2+}$.}
\label{ediag}
\end{figure}


\begin{thebibliography}{42}
\expandafter\ifx\csname natexlab\endcsname\relax\def\natexlab#1{#1}\fi
\expandafter\ifx\csname bibnamefont\endcsname\relax
  \def\bibnamefont#1{#1}\fi
\expandafter\ifx\csname bibfnamefont\endcsname\relax
  \def\bibfnamefont#1{#1}\fi
\expandafter\ifx\csname citenamefont\endcsname\relax
  \def\citenamefont#1{#1}\fi
\expandafter\ifx\csname url\endcsname\relax
  \def\url#1{\texttt{#1}}\fi
\expandafter\ifx\csname urlprefix\endcsname\relax\def\urlprefix{URL }\fi
\providecommand{\bibinfo}[2]{#2}
\providecommand{\eprint}[2][]{\url{#2}}

\bibitem[{\citenamefont{Fukui et~al.}(1952)\citenamefont{Fukui, Yonezawa, and
  Shingu}}]{fukjcp}
\bibinfo{author}{\bibfnamefont{K.}~\bibnamefont{Fukui}},
  \bibinfo{author}{\bibfnamefont{T.}~\bibnamefont{Yonezawa}}, \bibnamefont{and}
  \bibinfo{author}{\bibfnamefont{H.}~\bibnamefont{Shingu}},
  \bibinfo{journal}{J. Chem. Phys.} \textbf{\bibinfo{volume}{20}},
  \bibinfo{pages}{722} (\bibinfo{year}{1952}).

\bibitem[{\citenamefont{Fukui}(1982)}]{fuksci}
\bibinfo{author}{\bibfnamefont{K.}~\bibnamefont{Fukui}},
  \bibinfo{journal}{Science} \textbf{\bibinfo{volume}{218}},
  \bibinfo{pages}{4574} (\bibinfo{year}{1982}).

\bibitem[{\citenamefont{Parr and Yang}(1984)}]{parryang84}
\bibinfo{author}{\bibfnamefont{R.~G.} \bibnamefont{Parr}} \bibnamefont{and}
  \bibinfo{author}{\bibfnamefont{W.}~\bibnamefont{Yang}}, \bibinfo{journal}{J
  Am Chem Soc} \textbf{\bibinfo{volume}{106}}, \bibinfo{pages}{4049}
  (\bibinfo{year}{1984}).

\bibitem[{\citenamefont{Chattaraj et~al.}(1995)\citenamefont{Chattaraj,
  Cedillo, and Parr}}]{ccp95}
\bibinfo{author}{\bibfnamefont{P.~K.} \bibnamefont{Chattaraj}},
  \bibinfo{author}{\bibfnamefont{A.}~\bibnamefont{Cedillo}}, \bibnamefont{and}
  \bibinfo{author}{\bibfnamefont{R.~G.} \bibnamefont{Parr}},
  \bibinfo{journal}{J. Chem. Phys.} \textbf{\bibinfo{volume}{103}},
  \bibinfo{pages}{7645} (\bibinfo{year}{1995}).

\bibitem[{\citenamefont{Itskowitz and Berkowitz}(1997)}]{ib97}
\bibinfo{author}{\bibfnamefont{P.}~\bibnamefont{Itskowitz}} \bibnamefont{and}
  \bibinfo{author}{\bibfnamefont{M.~I.} \bibnamefont{Berkowitz}},
  \bibinfo{journal}{J. Phys. Chem. A} \textbf{\bibinfo{volume}{101}},
  \bibinfo{pages}{5687} (\bibinfo{year}{1997}).

\bibitem[{\citenamefont{Ayers and Parr}(2000)}]{ap99}
\bibinfo{author}{\bibfnamefont{P.~W.} \bibnamefont{Ayers}} \bibnamefont{and}
  \bibinfo{author}{\bibfnamefont{R.~G.} \bibnamefont{Parr}},
  \bibinfo{journal}{J. Am. Chem. Soc.} \textbf{\bibinfo{volume}{122}},
  \bibinfo{pages}{2010} (\bibinfo{year}{2000}).

\bibitem[{\citenamefont{Parr et~al.}(2000)\citenamefont{Parr, Yang, Ayers, and
  Levy}}]{payaayle}
\bibinfo{author}{\bibfnamefont{R.~G.} \bibnamefont{Parr}},
  \bibinfo{author}{\bibfnamefont{W.}~\bibnamefont{Yang}},
  \bibinfo{author}{\bibfnamefont{P.~W.} \bibnamefont{Ayers}}, \bibnamefont{and}
  \bibinfo{author}{\bibfnamefont{M.}~\bibnamefont{Levy}},
  \bibinfo{journal}{Theor. Chem. Acc.} \textbf{\bibinfo{volume}{103}},
  \bibinfo{pages}{353} (\bibinfo{year}{2000}).

\bibitem[{\citenamefont{Ayers et~al.}(2002)\citenamefont{Ayers, Morrison, and
  Roy}}]{amr02}
\bibinfo{author}{\bibfnamefont{P.~W.} \bibnamefont{Ayers}},
  \bibinfo{author}{\bibfnamefont{R.~C.} \bibnamefont{Morrison}},
  \bibnamefont{and} \bibinfo{author}{\bibfnamefont{R.~K.} \bibnamefont{Roy}},
  \bibinfo{journal}{J. Chem. Phys.} \textbf{\bibinfo{volume}{116}},
  \bibinfo{pages}{8731} (\bibinfo{year}{2002}).

\bibitem[{\citenamefont{Geerlings et~al.}(2003)\citenamefont{Geerlings, Proft,
  and Langenaeker}}]{gpl03}
\bibinfo{author}{\bibfnamefont{P.}~\bibnamefont{Geerlings}},
  \bibinfo{author}{\bibfnamefont{F.~D.} \bibnamefont{Proft}}, \bibnamefont{and}
  \bibinfo{author}{\bibfnamefont{W.}~\bibnamefont{Langenaeker}},
  \bibinfo{journal}{Chem. Rev.} \textbf{\bibinfo{volume}{103}},
  \bibinfo{pages}{1793} (\bibinfo{year}{2003}).

\bibitem[{\citenamefont{Geerlings and Proft}(2008)}]{gp08}
\bibinfo{author}{\bibfnamefont{P.}~\bibnamefont{Geerlings}} \bibnamefont{and}
  \bibinfo{author}{\bibfnamefont{F.~D.} \bibnamefont{Proft}},
  \bibinfo{journal}{Phys. Chem. Chem. Phys.} \textbf{\bibinfo{volume}{10}},
  \bibinfo{pages}{3028} (\bibinfo{year}{2008}).

\bibitem[{\citenamefont{Ayers and Parr}(2008{\natexlab{a}})}]{ap08}
\bibinfo{author}{\bibfnamefont{P.~W.} \bibnamefont{Ayers}} \bibnamefont{and}
  \bibinfo{author}{\bibfnamefont{R.~G.} \bibnamefont{Parr}},
  \bibinfo{journal}{J. Chem. Phys.} \textbf{\bibinfo{volume}{129}},
  \bibinfo{pages}{054111} (\bibinfo{year}{2008}{\natexlab{a}}).

\bibitem[{\citenamefont{Ghosh and Islam}(2010)}]{gi09}
\bibinfo{author}{\bibfnamefont{D.~C.} \bibnamefont{Ghosh}} \bibnamefont{and}
  \bibinfo{author}{\bibfnamefont{N.}~\bibnamefont{Islam}},
  \bibinfo{journal}{Int. J. Quantum Chem.} \textbf{\bibinfo{volume}{111}},
  \bibinfo{pages}{11} (\bibinfo{year}{2010}).

\bibitem[{\citenamefont{Parr and Yang}(1989)}]{parrbook}
\bibinfo{author}{\bibfnamefont{R.~G.} \bibnamefont{Parr}} \bibnamefont{and}
  \bibinfo{author}{\bibfnamefont{W.}~\bibnamefont{Yang}},
  \emph{\bibinfo{title}{Density-Functional Theory of Atoms and Molecules}}
  (\bibinfo{publisher}{Oxford University Press}, \bibinfo{address}{New York},
  \bibinfo{year}{1989}).

\bibitem[{\citenamefont{Hohenberg and Kohn}(1964)}]{hk}
\bibinfo{author}{\bibfnamefont{P.}~\bibnamefont{Hohenberg}} \bibnamefont{and}
  \bibinfo{author}{\bibfnamefont{W.}~\bibnamefont{Kohn}},
  \bibinfo{journal}{Phys. Rev.} \textbf{\bibinfo{volume}{B864}},
  \bibinfo{pages}{136} (\bibinfo{year}{1964}).

\bibitem[{\citenamefont{von Barth}(2004)}]{vBscripta}
\bibinfo{author}{\bibfnamefont{U.}~\bibnamefont{von Barth}},
  \bibinfo{journal}{Physica Scripta} \textbf{\bibinfo{volume}{T109}},
  \bibinfo{pages}{9} (\bibinfo{year}{2004}).

\bibitem[{\citenamefont{Kohn and Sham}(1965)}]{KS}
\bibinfo{author}{\bibfnamefont{W.}~\bibnamefont{Kohn}} \bibnamefont{and}
  \bibinfo{author}{\bibfnamefont{L.~J.} \bibnamefont{Sham}},
  \bibinfo{journal}{Phys. Rev.} \textbf{\bibinfo{volume}{140}},
  \bibinfo{pages}{A1133} (\bibinfo{year}{1965}).

\bibitem[{\citenamefont{Perdew}(1985)}]{perde}
\bibinfo{author}{\bibfnamefont{J.~P.} \bibnamefont{Perdew}}, in
  \emph{\bibinfo{booktitle}{Density Functional Methods in Physics}}, edited by
  \bibinfo{editor}{\bibfnamefont{R.~M.} \bibnamefont{Dreizler}}
  \bibnamefont{and}
  \bibinfo{editor}{\bibfnamefont{J.}~\bibnamefont{da~Providencia}}
  (\bibinfo{publisher}{Plenum}, \bibinfo{address}{New York},
  \bibinfo{year}{1985}).

\bibitem[{\citenamefont{Almbladh and von Barth}(1985)}]{ulfca}
\bibinfo{author}{\bibfnamefont{C.-O.} \bibnamefont{Almbladh}} \bibnamefont{and}
  \bibinfo{author}{\bibfnamefont{U.}~\bibnamefont{von Barth}}, in
  \emph{\bibinfo{booktitle}{Density Functional Methods in Physics}}, edited by
  \bibinfo{editor}{\bibfnamefont{R.~M.} \bibnamefont{Dreizler}}
  \bibnamefont{and}
  \bibinfo{editor}{\bibfnamefont{J.}~\bibnamefont{da~Providencia}}
  (\bibinfo{publisher}{Plenum}, \bibinfo{address}{New York},
  \bibinfo{year}{1985}).

\bibitem[{\citenamefont{Perdew et~al.}(1982)\citenamefont{Perdew, Parr, Levy,
  and Balduz}}]{pplb82}
\bibinfo{author}{\bibfnamefont{J.~P.} \bibnamefont{Perdew}},
  \bibinfo{author}{\bibfnamefont{R.~G.} \bibnamefont{Parr}},
  \bibinfo{author}{\bibfnamefont{M.}~\bibnamefont{Levy}}, \bibnamefont{and}
  \bibinfo{author}{\bibfnamefont{J.~L.} \bibnamefont{Balduz}},
  \bibinfo{journal}{Phys. Rev. Lett.} \textbf{\bibinfo{volume}{49}},
  \bibinfo{pages}{1691} (\bibinfo{year}{1982}).

\bibitem[{\citenamefont{Bartolotti and Ayers}(2005)}]{ba05}
\bibinfo{author}{\bibfnamefont{L.~J.} \bibnamefont{Bartolotti}}
  \bibnamefont{and} \bibinfo{author}{\bibfnamefont{P.~W.} \bibnamefont{Ayers}},
  \bibinfo{journal}{J. Phys. Chem. A} \textbf{\bibinfo{volume}{109}},
  \bibinfo{pages}{1146} (\bibinfo{year}{2005}).

\bibitem[{\citenamefont{Cohen and Wasserman}(2007)}]{cowassfound}
\bibinfo{author}{\bibfnamefont{M.~H.} \bibnamefont{Cohen}} \bibnamefont{and}
  \bibinfo{author}{\bibfnamefont{A.}~\bibnamefont{Wasserman}},
  \bibinfo{journal}{J. Phys. Chem. A} \textbf{\bibinfo{volume}{111}},
  \bibinfo{pages}{2229} (\bibinfo{year}{2007}).

\bibitem[{\citenamefont{Gori-Giorgi and Savin}(2009)}]{gorisav}
\bibinfo{author}{\bibfnamefont{P.}~\bibnamefont{Gori-Giorgi}} \bibnamefont{and}
  \bibinfo{author}{\bibfnamefont{A.}~\bibnamefont{Savin}},
  \bibinfo{journal}{Int. J. Quantum Chem.} \textbf{\bibinfo{volume}{109}},
  \bibinfo{pages}{2410} (\bibinfo{year}{2009}).

\bibitem[{\citenamefont{Sagvolden and Perdew}(2008)}]{sagper}
\bibinfo{author}{\bibfnamefont{E.}~\bibnamefont{Sagvolden}} \bibnamefont{and}
  \bibinfo{author}{\bibfnamefont{J.~P.} \bibnamefont{Perdew}},
  \bibinfo{journal}{Phys. Rev. A} \textbf{\bibinfo{volume}{77}},
  \bibinfo{pages}{012517} (\bibinfo{year}{2008}).

\bibitem[{\citenamefont{Ayers et~al.}(2009)\citenamefont{Ayers, Yang, and
  Bartolotti}}]{ayb}
\bibinfo{author}{\bibfnamefont{P.~W.} \bibnamefont{Ayers}},
  \bibinfo{author}{\bibfnamefont{W.}~\bibnamefont{Yang}}, \bibnamefont{and}
  \bibinfo{author}{\bibfnamefont{L.~J.} \bibnamefont{Bartolotti}}, in
  \emph{\bibinfo{booktitle}{Chemical Reactivity Theory: A Density Functional
  View}}, edited by \bibinfo{editor}{\bibfnamefont{P.~K.}
  \bibnamefont{Chattaraj}} (\bibinfo{publisher}{CRC Press/Taylor \& Francis},
  \bibinfo{year}{2009}).

\bibitem[{\citenamefont{Hellgren and Gross}(2012)}]{hg11}
\bibinfo{author}{\bibfnamefont{M.}~\bibnamefont{Hellgren}} \bibnamefont{and}
  \bibinfo{author}{\bibfnamefont{E.~K.~U.} \bibnamefont{Gross}},
  \bibinfo{journal}{In Press, ArXiv:1108.3100v2}  (\bibinfo{year}{2012}).

\bibitem[{\citenamefont{Cohen et~al.}(1994)\citenamefont{Cohen,
  Ganduglia-Pirovano, and Kudrnovsky}}]{copiku1}
\bibinfo{author}{\bibfnamefont{M.~H.} \bibnamefont{Cohen}},
  \bibinfo{author}{\bibfnamefont{M.~V.} \bibnamefont{Ganduglia-Pirovano}},
  \bibnamefont{and}
  \bibinfo{author}{\bibfnamefont{J.}~\bibnamefont{Kudrnovsky}},
  \bibinfo{journal}{J. Chem. Phys.} \textbf{\bibinfo{volume}{101}},
  \bibinfo{pages}{8988} (\bibinfo{year}{1994}).

\bibitem[{\citenamefont{Cohen et~al.}(1995)\citenamefont{Cohen,
  Ganduglia-Pirovano, and Kudrnovsky}}]{copiku2}
\bibinfo{author}{\bibfnamefont{M.~H.} \bibnamefont{Cohen}},
  \bibinfo{author}{\bibfnamefont{M.~V.} \bibnamefont{Ganduglia-Pirovano}},
  \bibnamefont{and}
  \bibinfo{author}{\bibfnamefont{J.}~\bibnamefont{Kudrnovsky}},
  \bibinfo{journal}{J. Chem. Phys.} \textbf{\bibinfo{volume}{103}},
  \bibinfo{pages}{3543} (\bibinfo{year}{1995}).

\bibitem[{\citenamefont{Ayers et~al.}(2007)\citenamefont{Ayers, Proft, Borgoo,
  and Geerlings}}]{apbg07}
\bibinfo{author}{\bibfnamefont{P.~W.} \bibnamefont{Ayers}},
  \bibinfo{author}{\bibfnamefont{F.~D.} \bibnamefont{Proft}},
  \bibinfo{author}{\bibfnamefont{A.}~\bibnamefont{Borgoo}}, \bibnamefont{and}
  \bibinfo{author}{\bibfnamefont{P.}~\bibnamefont{Geerlings}},
  \bibinfo{journal}{J. Chem. Phys.} \textbf{\bibinfo{volume}{126}},
  \bibinfo{pages}{224107} (\bibinfo{year}{2007}).

\bibitem[{\citenamefont{Sablon et~al.}(2007)\citenamefont{Sablon, Proft, Ayers,
  and Geerlings}}]{spag07}
\bibinfo{author}{\bibfnamefont{N.}~\bibnamefont{Sablon}},
  \bibinfo{author}{\bibfnamefont{F.~D.} \bibnamefont{Proft}},
  \bibinfo{author}{\bibfnamefont{P.~W.} \bibnamefont{Ayers}}, \bibnamefont{and}
  \bibinfo{author}{\bibfnamefont{P.}~\bibnamefont{Geerlings}},
  \bibinfo{journal}{J. Chem. Phys.} \textbf{\bibinfo{volume}{126}},
  \bibinfo{pages}{224108} (\bibinfo{year}{2007}).

\bibitem[{\citenamefont{Fievez et~al.}(2008)\citenamefont{Fievez, Sablon,
  Proft, Ayers, and Geerlings}}]{fspag08}
\bibinfo{author}{\bibfnamefont{T.}~\bibnamefont{Fievez}},
  \bibinfo{author}{\bibfnamefont{N.}~\bibnamefont{Sablon}},
  \bibinfo{author}{\bibfnamefont{F.~D.} \bibnamefont{Proft}},
  \bibinfo{author}{\bibfnamefont{P.~W.} \bibnamefont{Ayers}}, \bibnamefont{and}
  \bibinfo{author}{\bibfnamefont{P.}~\bibnamefont{Geerlings}},
  \bibinfo{journal}{J. Chem. Theor. Comp.} \textbf{\bibinfo{volume}{4}},
  \bibinfo{pages}{1065} (\bibinfo{year}{2008}).

\bibitem[{\citenamefont{Perdew and Levy}(1997)}]{pl97}
\bibinfo{author}{\bibfnamefont{J.~P.} \bibnamefont{Perdew}} \bibnamefont{and}
  \bibinfo{author}{\bibfnamefont{M.}~\bibnamefont{Levy}},
  \bibinfo{journal}{Phys. Rev. B} \textbf{\bibinfo{volume}{56}},
  \bibinfo{pages}{16021} (\bibinfo{year}{1997}).

\bibitem[{\citenamefont{Cohen et~al.}(2008)\citenamefont{Cohen, Mori-S\'anchez,
  and Yang}}]{cmsy08}
\bibinfo{author}{\bibfnamefont{A.}~\bibnamefont{Cohen}},
  \bibinfo{author}{\bibfnamefont{P.}~\bibnamefont{Mori-S\'anchez}},
  \bibnamefont{and} \bibinfo{author}{\bibfnamefont{W.}~\bibnamefont{Yang}},
  \bibinfo{journal}{Science} \textbf{\bibinfo{volume}{321}},
  \bibinfo{pages}{792} (\bibinfo{year}{2008}).

\bibitem[{\citenamefont{Mori-S\'anchez
  et~al.}(2009)\citenamefont{Mori-S\'anchez, Cohen, and Yang}}]{mscy09}
\bibinfo{author}{\bibfnamefont{P.}~\bibnamefont{Mori-S\'anchez}},
  \bibinfo{author}{\bibfnamefont{A.~J.} \bibnamefont{Cohen}}, \bibnamefont{and}
  \bibinfo{author}{\bibfnamefont{W.}~\bibnamefont{Yang}},
  \bibinfo{journal}{Phys. Rev. Lett.} \textbf{\bibinfo{volume}{102}},
  \bibinfo{pages}{066403} (\bibinfo{year}{2009}).

\bibitem[{fnd()}]{fndisc}
\bibinfo{note}{Notice that there are cases where $\partial E_\xc/\partial N$ is
  discontinuous, but where this discontinuity is exactly canceled by the
  discontinuity of the Fukui function (see e.g. the Hartree functional). In
  this case, $\D_\xc$ is thus zero.}

\bibitem[{\citenamefont{Krieger et~al.}(1992)\citenamefont{Krieger, Li, and
  Iafrate}}]{kli}
\bibinfo{author}{\bibfnamefont{J.~B.} \bibnamefont{Krieger}},
  \bibinfo{author}{\bibfnamefont{Y.}~\bibnamefont{Li}}, \bibnamefont{and}
  \bibinfo{author}{\bibfnamefont{G.~J.} \bibnamefont{Iafrate}},
  \bibinfo{journal}{Phys. Rev. A} \textbf{\bibinfo{volume}{45}},
  \bibinfo{pages}{101} (\bibinfo{year}{1992}).

\bibitem[{\citenamefont{Grabo et~al.}(2000)\citenamefont{Grabo, Kreibich,
  Kurth, and Gross}}]{gkkg}
\bibinfo{author}{\bibfnamefont{T.}~\bibnamefont{Grabo}},
  \bibinfo{author}{\bibfnamefont{T.}~\bibnamefont{Kreibich}},
  \bibinfo{author}{\bibfnamefont{S.}~\bibnamefont{Kurth}}, \bibnamefont{and}
  \bibinfo{author}{\bibfnamefont{E.~K.~U.} \bibnamefont{Gross}}, in
  \emph{\bibinfo{booktitle}{Strong Coulomb correlations in electronic structure
  calculations: beyond the Local Density Approximation}}, edited by
  \bibinfo{editor}{\bibfnamefont{V.}~\bibnamefont{Anisimov}}
  (\bibinfo{publisher}{Gordon and Breach}, \bibinfo{address}{Amsterdam},
  \bibinfo{year}{2000}).

\bibitem[{\citenamefont{Tempel et~al.}(2009)\citenamefont{Tempel, Mart√≠nez,
  and Maitra}}]{maitrasc}
\bibinfo{author}{\bibfnamefont{D.}~\bibnamefont{Tempel}},
  \bibinfo{author}{\bibfnamefont{T.~J.} \bibnamefont{Mart√≠nez}},
  \bibnamefont{and} \bibinfo{author}{\bibfnamefont{N.~T.}
  \bibnamefont{Maitra}}, \bibinfo{journal}{J. Chem. Theor. Comp.}
  \textbf{\bibinfo{volume}{5}}, \bibinfo{pages}{770} (\bibinfo{year}{2009}).

\bibitem[{\citenamefont{Helbig et~al.}(2009)\citenamefont{Helbig, Tokatly, and
  Rubio}}]{helsc}
\bibinfo{author}{\bibfnamefont{N.}~\bibnamefont{Helbig}},
  \bibinfo{author}{\bibfnamefont{I.~V.} \bibnamefont{Tokatly}},
  \bibnamefont{and} \bibinfo{author}{\bibfnamefont{A.}~\bibnamefont{Rubio}},
  \bibinfo{journal}{J. Chem. Phys.} \textbf{\bibinfo{volume}{131}},
  \bibinfo{pages}{224105} (\bibinfo{year}{2009}).

\bibitem[{\citenamefont{Balawender and Geerlings}(2005)}]{bg05}
\bibinfo{author}{\bibfnamefont{R.}~\bibnamefont{Balawender}} \bibnamefont{and}
  \bibinfo{author}{\bibfnamefont{P.}~\bibnamefont{Geerlings}},
  \bibinfo{journal}{J. Chem. Phys.} \textbf{\bibinfo{volume}{123}},
  \bibinfo{pages}{124103} (\bibinfo{year}{2005}).

\bibitem[{\citenamefont{Ayers and Parr}(2008{\natexlab{b}})}]{ap208}
\bibinfo{author}{\bibfnamefont{P.~W.} \bibnamefont{Ayers}} \bibnamefont{and}
  \bibinfo{author}{\bibfnamefont{R.~G.} \bibnamefont{Parr}},
  \bibinfo{journal}{J. Chem. Phys.} \textbf{\bibinfo{volume}{128}},
  \bibinfo{pages}{184108} (\bibinfo{year}{2008}{\natexlab{b}}).

\bibitem[{\citenamefont{Berkowitz and Parr}(1987)}]{bp87}
\bibinfo{author}{\bibfnamefont{M.}~\bibnamefont{Berkowitz}} \bibnamefont{and}
  \bibinfo{author}{\bibfnamefont{R.~G.} \bibnamefont{Parr}},
  \bibinfo{journal}{J. Chem. Phys.} \textbf{\bibinfo{volume}{88}},
  \bibinfo{pages}{2554} (\bibinfo{year}{1987}).

\bibitem[{\citenamefont{Chattaraj et~al.}(2007)\citenamefont{Chattaraj, Roy,
  Geerlings, and Sucarrat}}]{crgt07}
\bibinfo{author}{\bibfnamefont{P.}~\bibnamefont{Chattaraj}},
  \bibinfo{author}{\bibfnamefont{D.~R.} \bibnamefont{Roy}},
  \bibinfo{author}{\bibfnamefont{P.}~\bibnamefont{Geerlings}},
  \bibnamefont{and} \bibinfo{author}{\bibfnamefont{M.~T.}
  \bibnamefont{Sucarrat}}, \bibinfo{journal}{Theor. Chem. Acc.}
  \textbf{\bibinfo{volume}{118}}, \bibinfo{pages}{923} (\bibinfo{year}{2007}).

\end{thebibliography}
\end{document}